\documentclass[twocolumn,epjc3]{svjour3}
\usepackage{graphics}
\RequirePackage{graphicx}
\usepackage{amsmath}
\usepackage{amssymb}
\usepackage[draft]{hyperref}
\usepackage{textcomp}
\usepackage{multirow}
\begin{document}
%
\title{A detector module with highly efficient surface-alpha event rejection  operated in  CRESST-II Phase 2}
%
%
%

\author{R. Strauss \thanksref{1,2,e1} \and G. Angloher \thanksref{1} \and A. Bento \thanksref{3} \and C. Bucci \thanksref{4}\and L. Canonica \thanksref{4} \and A. Erb \thanksref{2,5} \and F.v.\,\,Feilitzsch \thanksref{2} \and N.\,\,Ferreiro \thanksref{1}\and P.\,\,Gorla \thanksref{4} \and A. G\"utlein \thanksref{2} \and D. Hauff \thanksref{1}  \and J. Jochum \thanksref{6} \and M. Kiefer \thanksref{1} \and H. Kluck \thanksref{8} \and H. Kraus \thanksref{7} \and J.-C. Lanfranchi \thanksref{2} \and J. Loebell \thanksref{6} \and A. M\"unster \thanksref{2} \and F.\,\,Petricca \thanksref{1} \and W. Potzel \thanksref{2} \and F. Pr\"obst \thanksref{1} \and F. Reindl \thanksref{1} \and S. Roth \thanksref{2} \and K. Rottler \thanksref{6} \and C. Sailer \thanksref{6} \and K. Sch\"affner \thanksref{4} \and J.\,\,Schieck \thanksref{8} \and S. Scholl \thanksref{6} \and S. Sch\"onert \thanksref{2} \and W. Seidel \thanksref{1} \and M.v.\,\,Sivers \thanksref{2,e2} \and  M. Stanger \thanksref{2} \and  L. Stodolsky \thanksref{1} \and C. Strandhagen \thanksref{6} \and A.\,\,Tanzke \thanksref{1} \and M. Uffinger \thanksref{6} \and A. Ulrich \thanksref{2} \and I. Usherov \thanksref{6} \and S. Wawoczny \thanksref{2} \and M. Willers \thanksref{2} \and M. W\"ustrich \thanksref{1} \and A. Z\"oller \thanksref{2} 
}

\thankstext{e1}{Corresponding author; e-mail: strauss@mpp.mpg.de}
\thankstext{e2}{Present address: Albert Einstein Center for Fundamental Physics, University of Bern, CH-3012 Bern, Switzerland}

\institute{Max-Planck-Institut f\"ur Physik,  D-80805 M\"unchen, Germany \label{1} \and Physik-Department, Technische Universit\"at M\"unchen \label{2}, D-85748 Garching, Germany   \and CIUC, Departamento de Fisica, Universidade de Coimbra, P3004 516 Coimbra, Portugal \label{3}\and INFN, Laboratori Nazionali del Gran Sasso, I-67010 Assergi, Italy \label{4} \and Walther-Mei\ss ner-Institut f\"ur Tieftemperaturforschung,  D-85748 Garching, Germany \label{5} \and Physikalisches Institut, Eberhard-Karls-Universit\"at T\"ubingen,   D-72076 T\"ubingen, Germany \label{6} \and Department of Physics, University of Oxford, Oxford OX1 3RH, United Kingdom \label{7} \and Institut f\"ur Hochenergiephysik der \"Osterreichischen Akademie der Wissenschaften, A-1050 Wien, Austria \linebreak and Atominstitut, Vienna University of Technology, A-1020 Wien, Austria\label{8} 
}


\date{Received: date / Revised version: date}
%

\maketitle
\abstract{
The cryogenic dark matter experiment\linebreak CRESST-II aims at the direct detection of WIMPs via elastic scattering off nuclei in scintillating CaWO$_4$ crystals. We present a new, highly improved, detector design installed in the current run of  CRESST-II Phase 2 with an efficient active rejection of surface-alpha backgrounds. Using CaWO$_4$ sticks instead of metal clamps to hold the target crystal, a detector housing with fully-scintillating inner surface could be realized. The presented detector  (TUM40) provides an excellent threshold of ${\sim}\,0.60$\,keV and a resolution of $\sigma\,{\approx}\,0.090$\,keV (at 2.60\,keV). With significantly reduced background levels, TUM40 sets stringent limits on the spin-independent WIMP-nucleon scattering cross section and probes a new region of parameter space for WIMP masses below 3\,GeV/c$^2$. In this paper, we discuss the novel detector design and the surface-alpha event rejection in detail. 
\PACS{
      {PACS-key}{discribing text of that key}   \and
      {PACS-key}{discribing text of that key}
     } 
} 

\authorrunning{R. Strauss, et al.}
\titlerunning{A novel detector module operated in CRESST-II Phase 2}

\section{Introduction}
There is compelling evidence that a significant fraction of the Universe is made of dark matter, suggesting new particles beyond the standard model \cite{Bertone:2004pz}. Weakly interacting massive particles (WIMPs) \cite{Jungman:1995df} are  well motivated candidates and might be detectable  with earth-bound experiments.

A variety of detectors were built during the last two decades searching for WIMP-induced nuclear recoils  with different target materials and techniques \cite{Cushman:2013zza}. While for higher WIMP masses ($m_\chi\gtrsim6\,$GeV/c$^2$) the liquid xenon based LUX \cite{Akerib:2013tjd} experiment reports the \linebreak strongest upper limit for the elastic spin-independent WIMP-nucleon cross-section ($7.6\cdot 10^{-10}$\,pb at $m_\chi\approx33$\,GeV/c$^2$), the cryogenic experiments SuperCDMS \cite{Agnese:2014aze} ($4.1\,$GeV/c$^2 \lesssim  m_\chi \lesssim 6.0\,$GeV/c$^2$), CDMSlite \cite{Agnese:2013jaa} \linebreak ($3.0\,$GeV/c$^2\lesssim m_\chi \lesssim 4.1\,$GeV/c$^2$) and CRESST-II \linebreak Phase 2 \cite{Angloher:2014myn} ($m_\chi\lesssim\,3.0\,$GeV/c$^2$) are  particularly sensitive to low-mass WIMPs.

Besides the long-standing claim for a WIMP detection by DAMA/LIBRA \cite{Bernabei:2010mq}, during the last years, also the experiments CoGeNT \cite{Aalseth:2012if}, CDMS-Si \cite{PhysRevLett.111.251301} and CRESST-II \cite{Angloher:2012vn} reported a signal excess which could possibly be interpreted as low-mass WIMPs. Under standard assumption of isoscalar WIMP-nucleon scattering \cite{Lewin:1995rx} these interpretations, however, are incompatible  with the strongest upper limits mentioned above. In addition, results from the XENON100 experiment \cite{Aprile:2012nq} and a re-analysis of the CRESST-II commissioning-run data \cite{PhysRevD.85.021301} disfavour a low-mass WIMP scenario for  cross-sections accessible by the respective experiments.  \\
Consequently, for  CRESST-II Phase 2 a new detector design was developed providing a strongly reduced background level. First data from CRESST-II Phase 2 \cite{Angloher:2014myn} acquired by a new detector module (called TUM40) presented here explore a new region of parameter space below $\sim$3\,GeV/c$^2$ and strongly constrain the results from CRESST-II. It rules out the WIMP solution at $m_\chi \approx 11.6$\,GeV/c$^2$ (M2) and disfavours the solution at $m_\chi \approx 25.3$\,GeV/c$^2$ (M1).

\section{The CRESST-II experiment}
\subsection{Experimental basics}\label{sec:CRESST}
The Cryogenic Rare Event Search with Superconducting Thermometers (CRESST) uses scintillating CaWO$_4$ crystals as target material \cite{Angloher:2012vn}. This unique multi-\linebreak element approach makes CRESST-II sensitive to a wide range of WIMP masses $m_\chi$. Due to the $A^2$-dependence of the coherent spin-independent WIMP-nucleon cross-section \cite{Jungman:1995df}, the expected event rate is strongly enhanced for scatters off the heavy element W ($A\approx184$) which - for the obtained threshold - is the dominant target nucleus for $m_\chi\gtrsim 5$\,GeV/c$^2$; for lower WIMP masses, however, mostly O ($A\approx16$) and Ca ($A\approx40$) scatters can be observed \cite{Angloher:2014myn}. The scintillation light output of CaWO$_4$   depends strongly on the ionization strength of the interacting particle \cite{birks1964theory}. This effect is called quenching and is utilized in CRESST-II detectors for particle discrimination \cite{Strauss:2014ab}.

In a CRESST-II detector module, a cylindrical \linebreak CaWO$_4$ crystal of typically 300\,g in mass and 4\,cm in diameter and height is equipped with a W transition-edge-sensor (TES) to measure phonons induced by a particle interaction. It is referred to as the phonon detector. In addition, a light detector which consists of a $\sim500\,\mu$m thick silicon on sapphire disc of 4\,cm in diameter equipped with a W-TES is used as an absorber for the scintillation light. Both detectors are read out simultaneously with SQUIDs.

In the conventional detector design \cite{Angloher:2012vn}, the CaWO$_4$ target crystal is held in place by bronze clamps  covered by a thin film of Al. The elasticity of bronze avoids stress due to different thermal contraction of the various detector components, which otherwise might induce phonon events (with no light signal associated) that could mimic recoils of heavy nuclei \cite{Astrom2006262}.

The phonon and light detectors are surrounded by a polymeric multilayer foil (commercially available under the label \textit{VM2002}). It is highly reflective improving the light collection. Furthermore, it is scintillating to establish an active veto against surface-alpha decays. Additional light produced by alpha particles hitting the foil is used to veto the corresponding nuclear recoils entering the target crystal. 


\subsection{CRESST-II Phase 2}\label{sec:upgrade}
In July 2013, a new dark matter run of CRESST-II  with upgraded detectors was started (Phase 2). A total of 18 modules,  corresponding to an overall target mass of ${\sim}5$\,kg of CaWO$_4$ were installed. Seventeen modules are fully operational, 11 of which are of the conventional detector design. To reduce the surface backgrounds observed in the previous run of CRESST-II \cite{Angloher:2012vn} which are suspected to originate from a $^{210}$Po contamination of the bronze clamps, material selection and Rn-prevention methods were improved. For the latter, Rn-depleted air, supplied by the CUORE group \cite{Artusa:2014lgv} was used during assembly and mounting of the detectors.

In addition, six alternative detector modules of three different new designs (two each), which aimed at providing an efficient rejection of surface-alpha events were installed: 1) the target crystal is held by CaWO$_4$ sticks (subject of this paper), 2) a smaller CaWO$_4$ crystal, called carrier, carrying the TES is held by  bronze \linebreak clamps\footnote{The clamps are covered with scintillating Parylene.} and is glued to the main target crystals (called ``carrier-type'') and 3) a carrier-type detector is surrounded by a cup-shaped Si light detector (called \linebreak ``beaker-type'') \cite{freindl:statusCRESST}. For the two latter designs, background identification relies on phonon pulse-shape discrimination between the carrier and the main crystal.

The neutron shielding of the CRESST-II setup was augmented by an innermost layer of 3.5\,cm of polyethylene to further moderate  neutrons. The first (${\sim}105$) live-days of  CRESST-II Phase 2  were used for a low-mass WIMP analysis \cite{Angloher:2014myn} and - in the present paper - for an investigation of the detector performance and backgrounds. Data recorded since January 2014 are still blinded for future analysis.\\

\subsection{Illustration of CRESST-II data}\label{sec:illustration}
The two-channel readout of CRESST-II detectors allows one  to derive  $E_p$ and $E_l$, the measured energies in the phonon and light detector, respectively. Both channels are calibrated with gamma sources (typically  $^{57}$Co). In the phonon detector,  an energy $E_p=122$\,keV is assigned to the signal of a 122\,keV gamma. The energy of the detected scintillation light of such an event is defined as $E_l=122\,$keV$_{ee}$, the so-called electron-equivalent energy\footnote{The unit [keV$_{ee}$] is equivalent to [keV] in the calculations below.}. Consequently,  the relative light output, called light yield $LY=E_l/E_p$, is defined as 1 for 122\,keV gammas.

The $LY$ (and to a lesser certain extent also $E_p$) depends on the type of particle interaction. Due to (light) quenching,  nuclear recoils have a reduced light output. This effect is quantified by  Quenching Factors (QF). Since the total energy is shared between phonons and emitted photons, nuclear recoils have a slightly ($\mathcal{O}$(5\%)) higher phonon signal compared to electron recoils of the same energy \cite{Angloher:2014myn}.  Considering this, the event-type independent total deposited energy\footnote{In \cite{Strauss:2014ab} the event-type independent  total deposited energy is denoted as $E_r$. The correction is performed, but not explicitly mentioned.}  $E$  can  be expressed as
\begin{equation}
E = \eta E_l +(1-\eta)E_p = E_p (1-\eta (1-LY))
\end{equation}
for an event of a certain $LY$. The parameter $\eta$ represents the fraction of the total energy deposition escaping the crystal as scintillation light for an event at $LY=1$. For the crystal TUM40  a value of $\eta=0.066\pm0.004$(stat.) was determined in-situ \cite{Angloher:2014myn} which is in agreement with independent measurements performed with other CaWO$_4$ crystals, e.g. using alpha-induced nuclear recoils for calibration \cite{kiefer2012phd}.\\ 
CRESST-II data is usually displayed in the $LY$ vs. $E$ plane where event populations of different types of particle interactions appear as distinct  horizontal bands. For CRESST-II detectors, the mean of the dominant electron recoil band $LY_{e}(E)$ is  parametrized empirically by 
\begin{equation}\label{equ:ly_description}
LY_{e}(E) = (p_0+p_1E) (1 - p_2\exp(-E/p_3)),
\end{equation}
where the parameters $p_0$, $p_1$, $p_2$ and $p_3$ are derived by a fit to the data\footnote{The values found for the crystal TUM40 are: $p_0=0.938$, $p_1\,{=}\,4.6\cdot 10^{-5}$\,keV$^{-1}$, $p_2=0.389$ and $p_3=19.34$\,keV.} \cite{Strauss:2014ab}. The first factor in equation \ref{equ:ly_description} depends on the calibration of the detector. Due to the choice of the $LY$ normalization the parameter $p_0$ is usually very close to 1. The parameter $p_1$ which accounts for linear corrections to the calibration is in most cases consistent with zero and therefore neglected in the following. The second factor in equation \ref{equ:ly_description} describes the exponential reduction of the $LY$ towards lower $E$ which is known as non-proportionality effect \cite{2008ITNS...55.1049M,Lang:2009uh,sabine_phD}.   As phenomenologically explained by Birks' law, the relative scintillation light output is reduced for low-energy electrons due to their increased local energy loss \cite{birks1964theory}. The reduction of the LY for a certain kind of particle interaction $x=e,\alpha,O,Ca,W$ is described by $QF_x$ which we define as
\begin{equation}\label{equ:qf_definition}
QF_x(E)=\frac{LY_x(E)}{p_0}.
\end{equation}
For electron recoils, this effect  depends on the individual  CaWO$_4$ crystal used and is described by (see equation \ref{equ:ly_description} and \ref{equ:qf_definition}): $QF_e(E)=1 - p_2\exp(-E/p_3)$.\\
$QF_x$ for alpha particles and nuclear recoils is, in general, energy dependent \cite{Strauss:2014ab,sabine_phD}. However, in the  region-of-interest (ROI) for the dark matter search ($E\lesssim40$\,keV) constant QFs are a sufficient approximation\footnote{The values of the QFs depend, to a certain extent, also on the individual crystal (e.g. due to different optical properties). In \cite{Strauss:2014ab} a model is presented to account for this effect. The values given here are mean values measured in CRESST-II detectors.}:    $QF_\alpha=22$\% \cite{Angloher:2012vn}, $QF_O=(11.2{\pm}0.5)$\%, $QF_{Ca}=(5.94{\pm}0.49)$\% and $QF_{W}=(1.72{\pm}0.21)$\% \cite{Strauss:2014ab}.\\ 
Due to statistical fluctuations of the phonon and light signals the event bands have a finite width $\sigma$  which can  be well described by a Gaussian \cite{Strauss:2014ab}. The width $\sigma$ (in $LY$) of a certain recoil band is given by 
\begin{equation}
\sigma_x(E) = \frac{1}{E}\sqrt{\sigma_{l}^2+(\frac{dE_l}{dE} \sigma_{ph})^2+S_1  E_l +S_2 E_l^2},
\end{equation}
with $\sigma_{l}$ and $\sigma_{ph}$ being the baseline fluctuation of phonon and light detector, respectively,  $S_1$ the deposited energy  per detected photon (accounting for photon statistics) and $S_2$ accounting for possible position dependencies in light production and detection (not relevant in the ROI)\footnote{For TUM40 these parameters are $\sigma_{l}=269$\,eV, $\sigma_{ph}=91$\,eV, $S_1=256$\,eV and $S_2\approx 0$.}.

\begin{figure}
\centering
\includegraphics[width=0.5\textwidth]{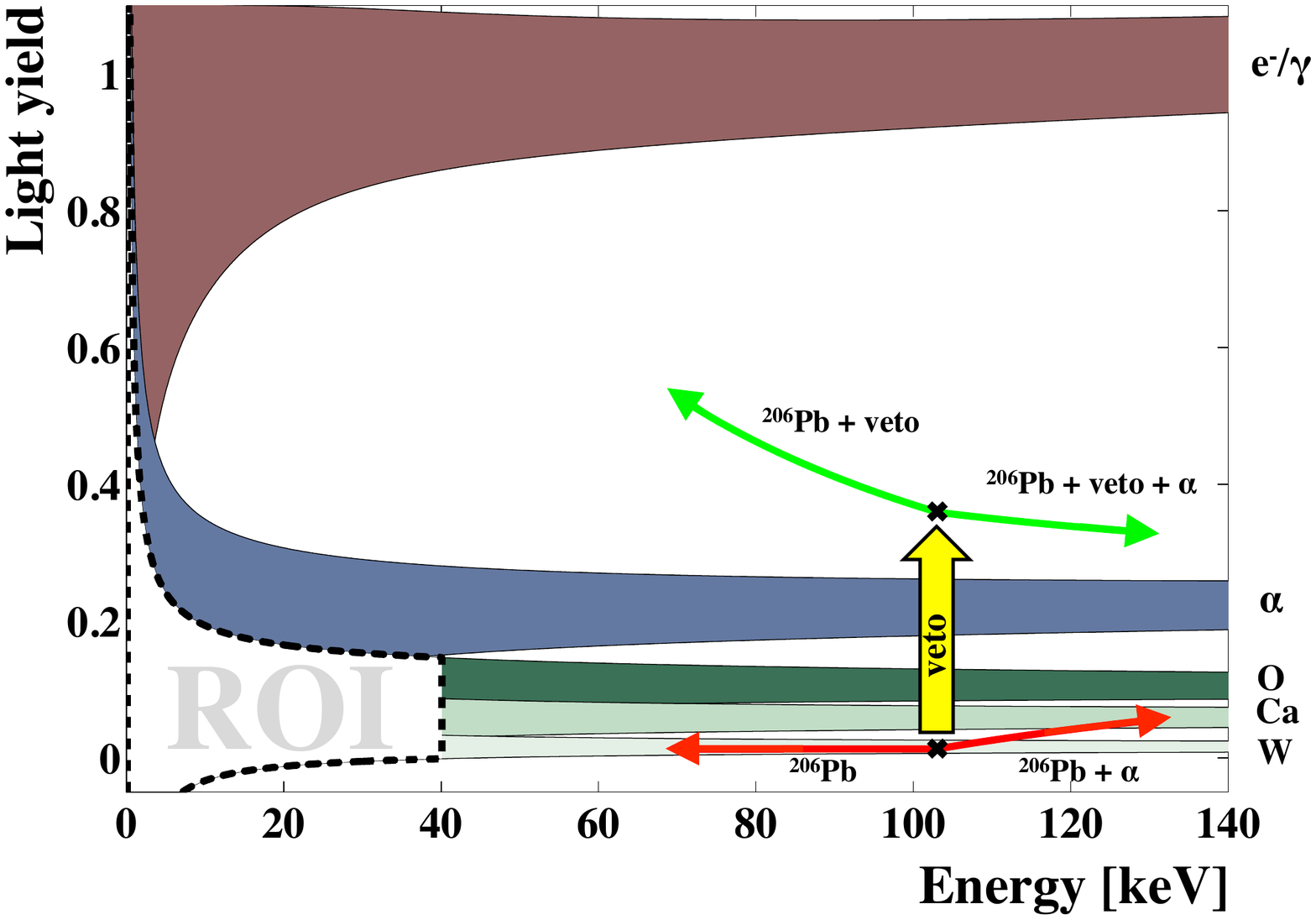}
\caption{Schematic plot of light yield  vs. energy. Separate horizontal event bands arise:  of beta/gamma events  at $LY\sim$1 (decreasing in $LY$ at energies $\lesssim20$\,keV, see equation \ref{equ:ly_description}),  of (degraded) alpha events at $LY\sim$0.22 and of nuclear recoils off O, Ca and W at $LY\lesssim$0.1. Possible surface $^{206}$Pb events appear at $\sim$103\,keV (at $LY\sim0.01$) and - depending on their origin - above or below this energy (red arrows). If the corresponding alpha particles hit the scintillating housing, additional light is produced (veto) which shifts these events out of the nuclear-recoil bands (green arrows). The region of interest  for dark matter search is indicated by a dashed black line. It includes the nuclear-recoil bands (O, Ca and W) and extends in energy from threshold (here: 0.6\,keV) to 40\,keV. }
\label{fig:LYcartoon}       
\end{figure}

In Fig. \ref{fig:LYcartoon}, where the LY is plotted against the (event-type independent) total deposited energy $E$, a schematic view of the event bands (beta/gamma, alpha, O, Ca and W) is shown. Due to the finite resolution these populations overlap at lower energies. The ROI for WIMP search which includes all three nuclear-recoil bands (O, Ca and W) and extends in $E$ from threshold (here: 0.6\,keV) to 40\,keV  is depicted with dashed black lines. 

\subsection{Observed background in CRESST-II}\label{observedBackgrounds}
During the previous dark matter run of CRESST-II,  730\,kg-days of exposure have been acquired by 8 detector modules with a total target mass of $\sim\,2.4$\,kg \cite{Angloher:2012vn}. Several types of backgrounds are  identified, mainly originating  from intrinsic contaminations of the CaWO$_4$ crystals and from contaminations of the direct vicinity of the detectors.  In the electron-recoil band a variety of beta-spectra and  gamma-lines up to the MeV range are visible in typical CRESST-II detectors \cite{Lang:2009uh}. At low energies $E<100$\,keV the spectra  of  commercially available crystals are usually dominated by intrinsic contamination of $^{227}$Ac and $^{210}$Pb \cite{Lang:electron_background}. Their mean count rate in the ROI ranges from 6 to 30/(kg\,keV\,day). For crystals from in-house production at the Techni\-sche Universit\"at M\"unchen (TUM) the background level could be reduced significantly to 3.44/(kg\,keV\,day) in the ROI (for TUM40) \cite{strauss:2014part2}. Peaks originating from cosmogenic activation of different tungsten isotopes (electron-capture transitions), dominate the low-energy spectrum \cite{strauss:2014part2}.

\begin{figure}
\centering
\includegraphics[width=0.4\textwidth]{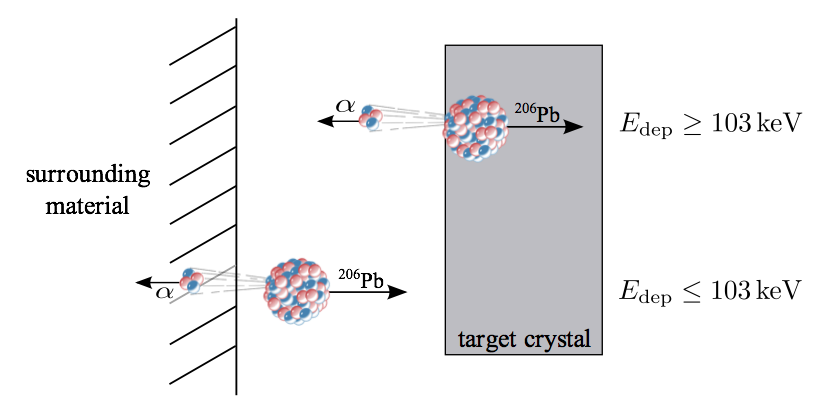}
\caption{Illustration of $^{210}$Po decays occurring either in surrounding material or close to the surface of the target crystal \cite{Angloher:2012vn}. }
\label{fig:PbBackground}       
\end{figure}

In the alpha band, several populations of events were observed. The distinct alpha peaks from intrinsic contamination by natural decay chains, by rare earth metals (e.g. $^{147}$Sm, $^{144}$Nd) and by the radioactive isotope $^{180}$W \cite{Cozzini:2004vd} have  energies in the MeV range and are far off the ROI \cite{muenster_2014,strauss:2014part2}. However, if alpha  emitters are embedded in the bulk of material surrounding the detectors (e.g. the bronze clamps) the corresponding alpha particles can loose part of their energy before being absorbed in the CaWO$_4$ crystal. This population of  degraded alphas can leak down to the ROI \cite{Angloher:2012vn}. Recoiling nuclei from surface alpha contamination, either on the crystal or on surrounding materials, show up at very low $LY$ comparable with that of W recoils. For CRESST-II detectors mainly $^{206}$Pb recoils  from $^{210}$Po decays (see Fig. \ref{fig:PbBackground})  are relevant:
\begin{equation}
^{210}\mathrm{Po} \rightarrow ^{206}\mathrm{Pb} (103\,\mathrm{keV}) + \alpha (5.3\,\mathrm{MeV}).
\end{equation}
If the decay occurs sufficiently close to surfaces the full  energy of 103\,keV is detected. However, Pb-recoils from $^{210}$Po implanted in surrounding surfaces, deposit less energy in the phonon detector. In contrast, for Pb-events originating from the crystal's surface part of the alpha energy is deposited additionally in the target crystal. The different classes of $^{206}$Pb recoils observed at low $LY$ ($QF_{Pb}\sim1.4$\,\%) are illustrated as red arrows in Fig. \ref{fig:LYcartoon}.

\section{The novel detector design}\label{sec:newPrototype}

\subsection{Fully-scintillating detector housing}\label{sec:fullyScintHouse}
In the previous dark matter run of CRESST-II, surface-alpha decays of $^{210}$Po caused the highest identified background contribution. A maximum-likelihood analysis attributed up to ${\sim}25$\,\% of the excess events \cite{Angloher:2012vn} to $^{206}$Pb recoils from a contamination on the non-\linebreak scintillating bronze clamps holding the CaWO$_4$ crystal. Pb-recoils which occur at $LY$s as low as W-recoils can be shifted out of the nuclear-recoil region if the corresponding alpha produces additional light. This can be achieved by surrounding the crystal   by scintillating material completely.  It was reported earlier \cite{Angloher:2012vn,Angloher2009270} that  the polymeric foil surrounding the target crystal scintillates sufficiently  to establish such a veto. However, all attempts to cover the  bronze clamps with scintillating material (e.g. plastic scintillator) have failed. Thermo-mechanical stress in the clamp-plastic bilayer can relax and cause false signals in the phonon detector. Since pulse-shape differences are not sufficient to fully discriminate such events and no scintillation light is produced, at low energies, such events can mimic recoils of heavy nuclei.

The new detector design presented here uses a different approach to avoid any non-scintillating surface inside the detector housing and, thus, to reach an efficient surface-event rejection. CaWO$_4$  sticks, penetrating the polymeric foil through a set of tightly fitting holes and spring-loaded by pure bronze clamps on the outside of the Cu-housing, hold the CaWO$_4$ crystal in place. Thereby, elasticity of the system is maintained down to mK temperatures while solely scintillating material is in line-of-sight of the CaWO$_4$ target crystal.

\begin{figure}
\centering
\includegraphics[width=0.45\textwidth]{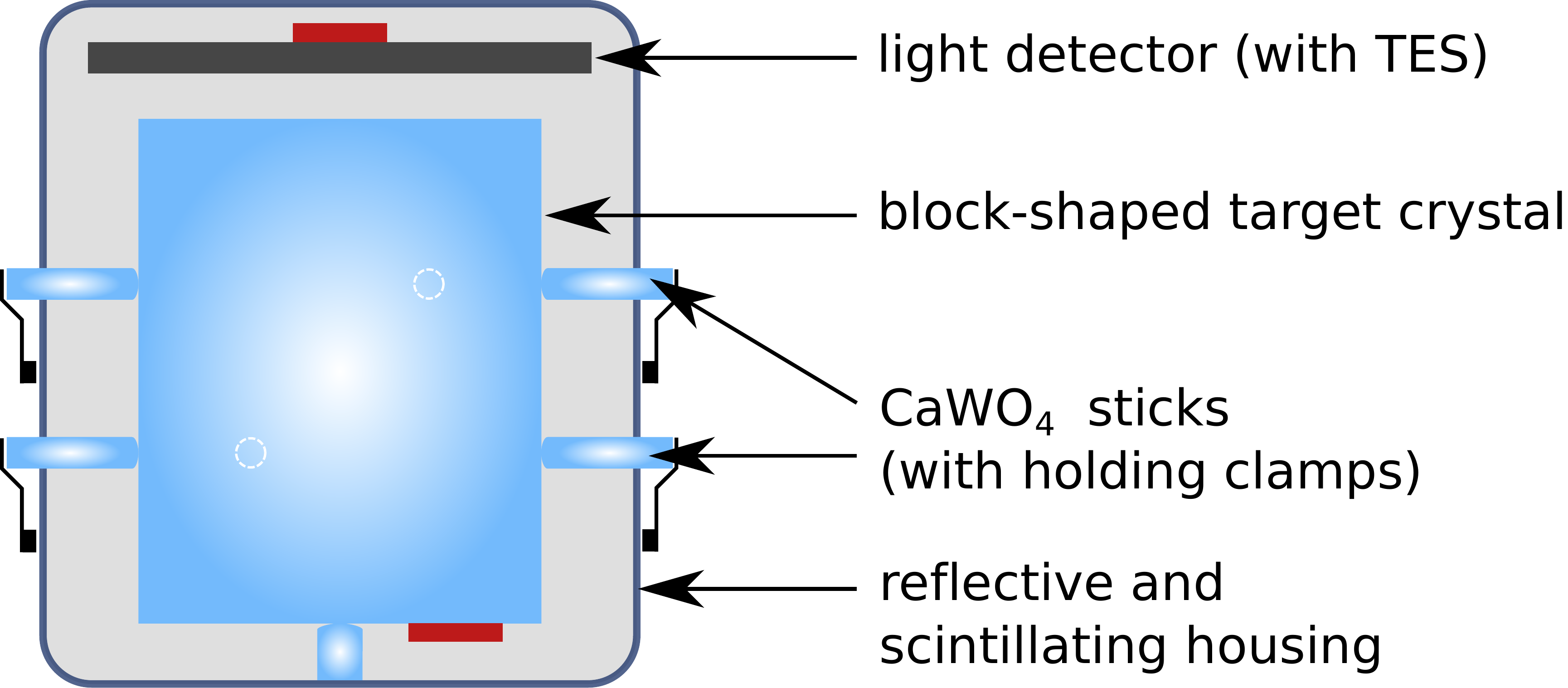}
\caption{Schematic view of the novel fully-scintillating detector module. A block-shaped target crystal  is held by CaWO$_4$ sticks. The bottom stick  is fixed in the copper support structure and the eight side sticks  are held by bronze clamps from outside the holder. The positions of the sticks are indicated by white circles.  Together with the scintillating polymeric foil an active veto against all surface events is realized.  }
\label{fig:stickHolder}       
\end{figure}

Fig. \ref{fig:stickHolder} shows a sketch of the novel detector design. A 249\,g block-shaped (rectangular) CaWO$_4$ crystal of 32\,mm edge length and 40\,mm height is held by eight CaWO$_4$ sticks  from the side  (diameter 2.5\,mm, length 7.5\,mm) and one CaWO$_4$ stick from the bottom (diameter 4.0\,mm, length 8.2\,mm). The CaWO$_4$ crystal and the sticks are produced at the TUM (see section \ref{sec:crystal}). 	The side sticks are pressed  onto the crystal by bronze clamps and the bottom stick is fixed in the copper support structure.  The sticks have a polished spherical end which provides a point-like contact with the target crystal. The remaining surface is covered with polymeric foil (VM2002). A conventional silicon on sapphire light detector is mounted onto the novel detector holder.

With this design, each nuclear recoil from  alpha decays occurring on any surface in the holder is vetoed by additional scintillation light from the alpha. Two different cases have to be considered:\\
\begin{enumerate}
\item  $^{210}$Po decays on the surface of the target crystal deposit (in the phonon detector) the full energy of the $^{206}$Pb recoil (103\,keV) plus potentially part of the energy  of the corresponding alpha particle  depending on the implantation depth\footnote{Decays of $^{222}$Rn and its daughters impact the respective nuclei into the CaWO$_4$ material. Consequently, the $^{210}$Po nuclei end up at different (implantation) depths  of $\mathcal{O}$(100\,nm) with respect to the surface.}. In the light detector, the scintillation light of the Pb-recoil and of part of the alpha energy, as well as the additional light produced in the foil or CaWO$_4$ are measured. The energy corresponding to the additional light detected is called $E_{veto}$. The LY of this kind of surface event  $LY_s$ can be expressed as
\begin{equation}\label{equ:ly_surface_1}
LY_{s} = p_0QF_{Pb}+\frac{E_{veto}+p_0QF_{\alpha}(E-103\,\text{keV})}{E},
\end{equation}
with $QF_{Pb}\sim1.4$\,\% \cite{Angloher:2012vn}.

\item $^{210}$Po decays on the surface of the polymeric foil or the CaWO$_4$ sticks induce Pb-recoils in the CaWO$_4$ crystal which deposit 103\,keV or less. In the light channel, the light of the Pb-recoil plus the scintillation light from foil or stick are detected. This gives a $LY$ of
 \begin{equation}\label{equ:ly_surface_2}
LY_{s} = p_0QF_{Pb}+\frac{E_{veto}}{E}. 
\end{equation}  
\end{enumerate}

Typical values of $E_{veto}$ for 5\,MeV alphas are \linebreak $\sim40\,$keV$_{ee}$ for the foil and  $\sim1000\,$keV$_{ee}$ for the CaWO$_4$ sticks. In Fig. \ref{fig:LYcartoon} the mean of the band of vetoed surface events is illustrated (green arrows), clearly separated from the nuclear-recoil bands by  additional light produced in the foil (yellow arrow).

In case a Pb-event is vetoed by the corresponding alpha interacting in a  CaWO$_4$ stick, $E_{veto}$ is sufficiently high to shift it to far above the electron-recoil band.  In addition, a certain fraction ($\sim1$\%) of the energy deposited by an alpha in a stick is visible in the phonon channel (see section \ref{sec:detectorPerformance}). This effect is neglected in equation \ref{equ:ly_surface_1} and \ref{equ:ly_surface_2}.      	  

\subsection{CaWO$_4$ crystals produced at TUM}\label{sec:crystal}
CaWO$_4$ crystals produced at the TUM are operated in the second phase of CRESST-II. A Czochralski growth furnace was set up with the intention to improve the radiopurity, the optical properties and to secure the supply of CaWO$_4$ single crystals for CRESST and future rare-event searches \cite{erb}. The crystals, which can be grown in a reproducible way and of sufficient size (diameter $>35$\,mm,  height $\gtrsim40\,$mm) since 2012, are further machined at the crystal lab of the TUM. Several techniques to investigate and improve the optical and scintillation properties have been developed \cite{Sivers:2012fd}. First measurements, in which in-house produced CaWO$_4$ crystals were operated as cryogenic detectors, were carried out at the CRESST test cryostat (located underground at the LNGS in Italy) to investigate radiopurity and light output \cite{strauss_PhD}. Using these experimental data, dedicated studies of the intrinsic alpha contamination \cite{muenster_2014,strauss_PhD} gave the first indication that the radiopurity (total internal alpha background $\sim3$\,mBq/kg) is improved by a factor of 2-10 with respect to commercially available crystals. Concerning scintillation and optical properties, the performance of crystals from external suppliers could not yet be accomplished. With a standard cylindrical CRESST-size crystal (TUM27), of which all sides are optically polished except the side facing the light detector, the fraction $L$ of the total deposited energy  detected as light (for a beta/gamma event) is $L{\sim}1.1$\% \cite{strauss_PhD}. This compares with a maximum of $L{\sim}2.4$\% achieved with the best commercial crystals \cite{kiefer2012phd}. Within these studies, the influence of roughening a larger part of the surface was measured. The light output of TUM27 was increased by 20\,\% to $L{\sim}1.3$\,\%, when the lateral surface was roughened in addition, while no influence on the phonon properties was observed.
   
\begin{figure}
\centering
\includegraphics[width=0.4\textwidth]{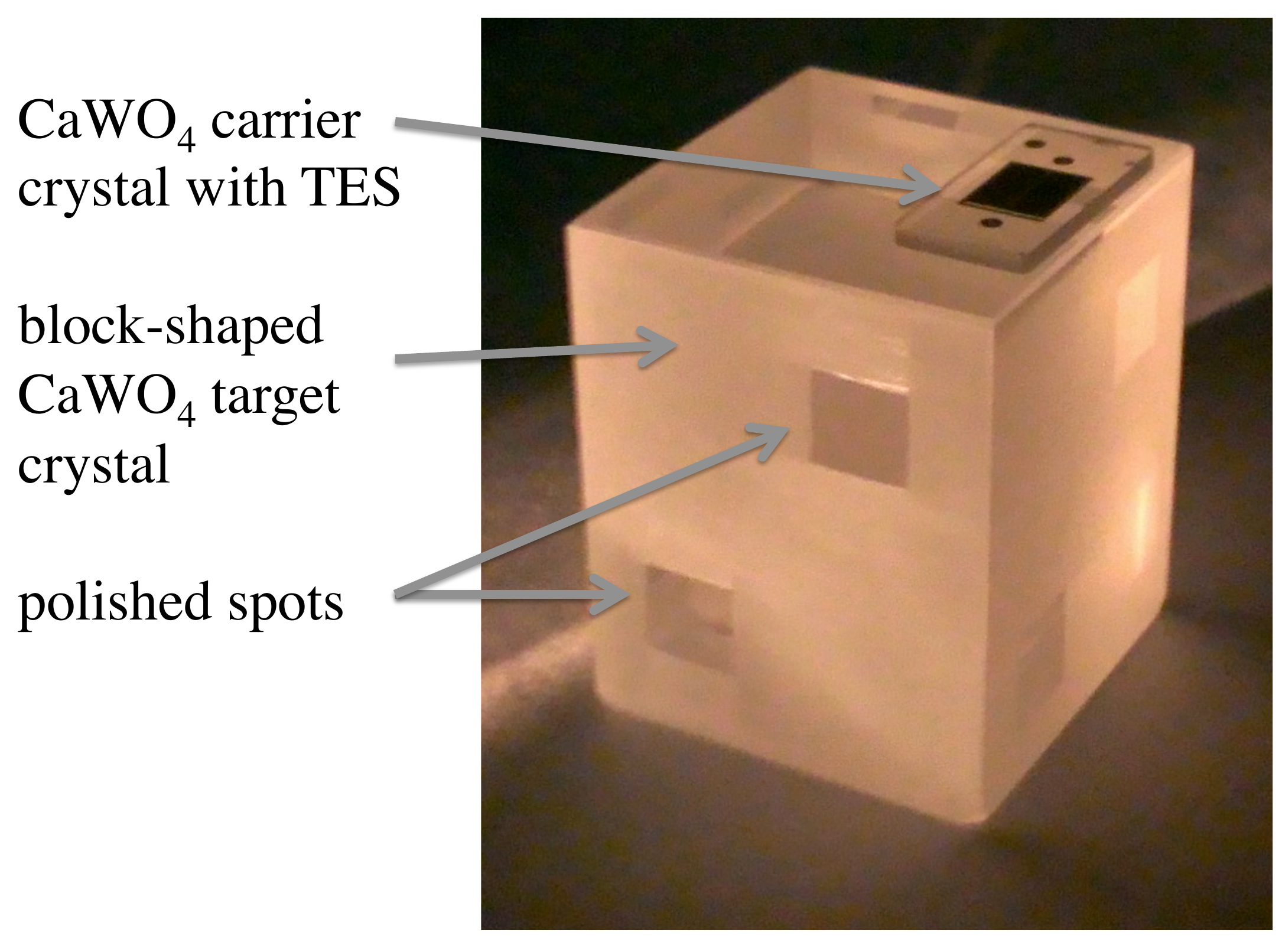}
\caption{The final state of the CaWO$_4$ crystal TUM40, as used in the experiment. The crystal was produced in-house at the TUM. A CaWO$_4$ carrier (10x20x2\,mm$^3$) with the W-TES is glued onto the polished side of the main crystal. The areas where the CaWO$_4$ sticks touch (7x7\,mm$^2$) remain polished while the rest of the crystal is roughened.}
\label{fig:pictureTUM40}       
\end{figure}

\noindent The block-shaped crystal TUM40,  which is supported by CaWO$_4$ sticks, was operated in several measurement campaigns at the test cryostat at LNGS before being installed into the CRESST setup. Even though the quality of the CaWO$_4$ material is comparable  to that of TUM27 (same raw materials and growth procedure) a  higher amount of detected light ($L{\sim}1.6$\,\%) was measured with this novel detector. Roughening five surfaces instead of only one did not significantly change $L$, however,  reduced position dependencies  of the light signal detected \cite{strauss_PhD}. Problems with stress relaxations occurred whenever CaWO$_4$ sticks were in contact with roughened surfaces \cite{strauss_PhD}. Therefore, the very spots (7\,mm x 7\,mm) where the sticks touch the crystal are left polished while the rest of the surface is roughened. The side facing the light detector is completely roughened while the opposite one is polished (in contact with bottom stick). A picture of the crystal TUM40 is shown in Fig. \ref{fig:pictureTUM40}.  
The TES is evaporated onto a separate CaWO$_4$ carrier (10x20x2\,mm$^3$) which is then glued onto the main absorber. This avoids having to expose the main crystal to very high temperatures (up to 600$^\circ$\,C) during the evaporation process of the TES which  degrades the scintillation light output \cite{Kiefer:2009pf}. The W-TES has a transition temperature of $T_c=20.2$\,mK. The carrier is glued onto the polished surface which is opposite to the light detector. The bond wires used to electrically connect the W-TES are fed through small slits in the surrounding polymeric foil \cite{strauss_PhD}.

\section{Results}
\subsection{Detector performance}\label{sec:detectorPerformance}
After a commissioning period of $\sim2$\,weeks, TUM40 and the light detector were operated under stable conditions in the CRESST-II setup. Before starting the dark matter run, a gamma-calibration campaign with a $^{57}$Co source was performed. The 122\,keV gamma-line is used for calibrating the heater pulses which are periodically injected onto the TESs. Heater pulses ensure  the calibration down to lowest energies and the long-term stability of the operation point  (see, e.g., \cite{Angloher:2012vn}).

\begin{figure}
\centering
\includegraphics[width=0.5\textwidth]{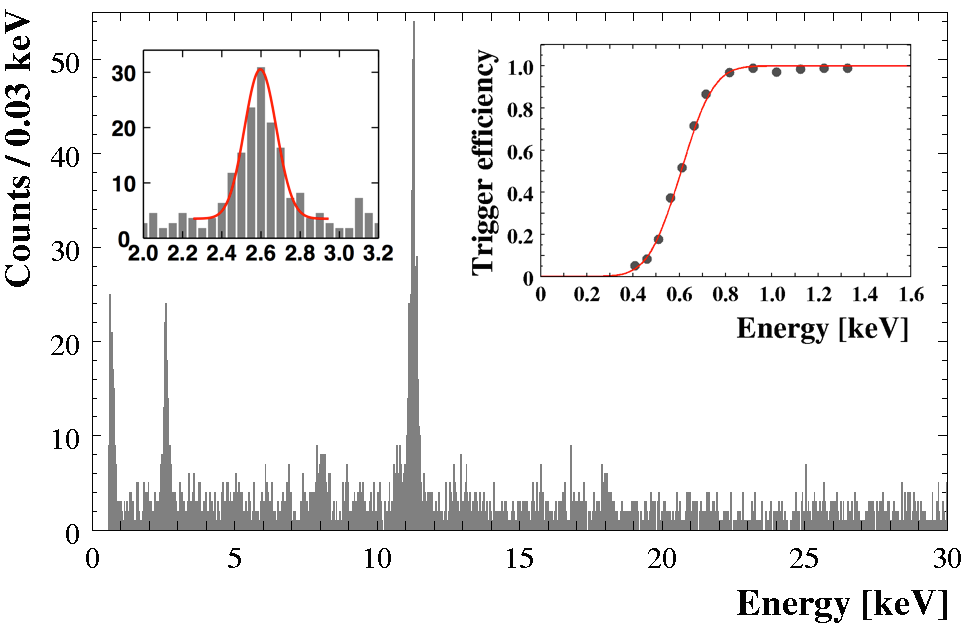}
\caption{Low-energy spectrum acquired with the module TUM40 and an exposure of 29\,kg-days in CRESST-II Phase 2. The prominent peaks at 11.74\,keV and 2.60\,keV originate from cosmogenic activation of $^{182}$W (see \cite{strauss:2014part2} for details). Left inset: Fit to the 2.60\,keV line (red line). A resolution of $\sigma{=}(0.090{\pm}0.010)\,$keV  is achieved \cite{Angloher:2014myn}. Right inset: T-he energy-dependent trigger efficiency of TUM40 which is measured with injected heater pulses (grey dots) is shown. A fit with an error function (red line) yields an energy threshold (50\% efficiency) of $E_{th}\approx(0.603\pm0.02)$\,keV \cite{Angloher:2014myn}.    }
\label{fig:histogram}       
\end{figure}

The use of the CaWO$_4$ sticks does not influence the detector performance, and no signs of microphonic noise are observed. Pulses from the phonon channel are comparable to the ones of conventional detector modules\footnote{Within the model for cryogenic particle detectors \cite{Probst:1995fk} decay times of thermal and non-thermal signal components of the particle pulses are $\tau_t{=}91.6$\,ms and $\tau_n{=}21.1$\,ms, respectively.}. An excellent resolution of $\sigma{=}(0.090{\pm}0.010)\,$keV (at  2.60\,keV) is achieved with TUM40 as shown in Fig. \ref{fig:histogram}. In the low-energy spectrum recorded in CRESST-II Phase 2 from an exposure of 29\,kg-days,  peaks from cosmogenic activation of tungsten isotopes are visible (see \cite{strauss:2014part2}). The trigger efficiency of the phonon detector was probed with injected heater pulses (grey dots in Fig. \ref{fig:histogram}, inset).  A fit with an error function (red line) gives a trigger threshold (50\% efficiency)  of $E_{th}\approx(0.603\pm0.02)$\,keV \cite{Angloher:2014myn}.

At an early stage of the CRESST experiment, events originating from mechanical stress relaxations were observed.  They were caused by a tight and rigid clamping of crystals \cite{Astrom2006262}. Dedicated studies \cite{strauss_PhD} and first data of CRESST-II Phase 2 show that  by using flexible bronze clamps mechanical stress on the stick-crystal interfaces is mitigated sufficiently with the novel detector design.

Particle events in the CaWO$_4$ sticks themselves produce scintillation light measured in the light detector. Due to the much smaller size of the sticks compared to the bulk crystal the relative light output of stick events is higher by a factor of up to three. To a certain extent, the induced phonon signal is detected in the TES of the main crystal through the stick-crystal interface. Since that connection is point-like only a degraded phonon signal from particle events in the sticks is expected. A dedicated measurement was performed at the test cryostat to quantify this so-called phonon-quenching effect at such interfaces \cite{strauss_PhD}.  A $^{147}$Sm alpha source  was placed at the CaWO$_4$-stick (bottom)  to calibrate the energy scale of particle events. A degradation of the phonon signal from events in the sticks by about two orders of magnitude (by a factor of $96{\pm}6$) was found. Due to the increased relative light output, the  population of beta/gamma and alpha events in the CaWO$_4$-sticks exhibits  light yields of 20-30 and 4-6, respectively, far off the bulk crystal's recoil bands. These event classes can be well separated from the ROI by LY (e.g. beta/gamma events with $7.7\sigma$ C.L. at $E=10\,$keV \cite{strauss_PhD}). In addition, stick events can be discriminated by pulse shape (the thermal component $\tau_t$ \cite{Probst:1995fk} is enhanced) at energies $\gtrsim10$\,keV. However, nuclear recoils occurring on the sticks' surfaces and the corresponding alpha particle hitting a non-scintillating surface (e.g. the bronze clamps outside the housing) might leak into the ROI at smallest energies $\lesssim1\,$keV. Considering the area of unvetoed surfaces, only $0.5\pm0.2$ events are expected from this background compared to $\mathcal{O}$(100) events of leakage from the beta/gamma band \cite{strauss:2014part2}. For the next-generation CRESST detectors, an additional scintillation veto is planned.

\subsection{Surface backgrounds}\label{sec:surface_backgrounds}
The main purpose of the novel detector design was to reduce significantly backgrounds related to surface-alpha decays. In CRESST-II  a mean rate of \linebreak $\sim0.05$/[kg\,day] from degraded alphas and Pb recoils was observed in the ROI ($\sim$12-40\,keV) of the eight detector modules operated \cite{Angloher:2012vn}.

\begin{figure}
\centering
\includegraphics[width=0.5\textwidth]{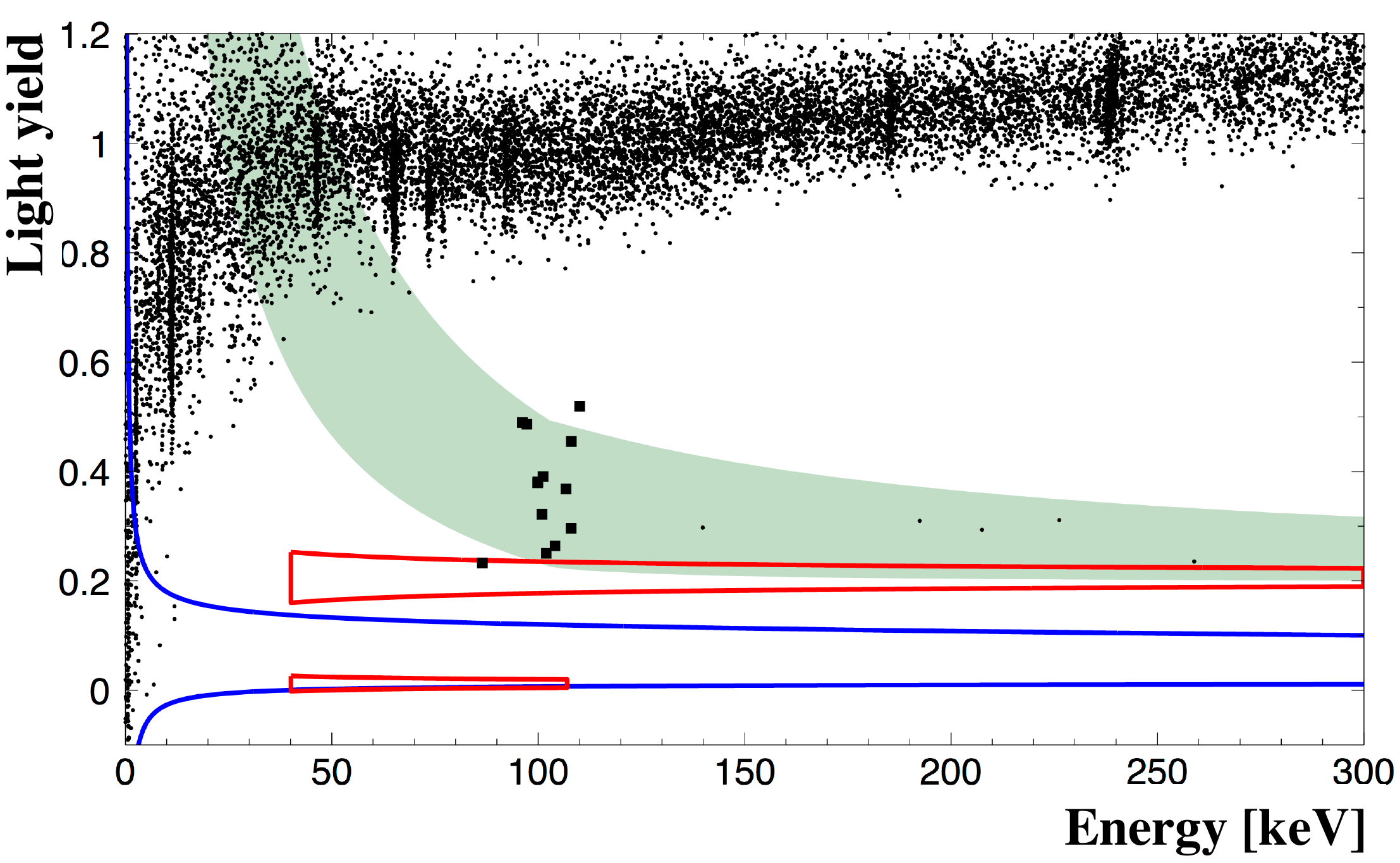}
\caption{Light yield (LY) vs. energy ($E$) of TUM40 data with an exposure of 29\,kg-days. The populated beta/gamma band is visible at $LY\sim1$. At $LY\sim0.1$, the 90\,\% upper bound of O recoils and the 90\,\% lower bound for W recoils are shown (full blue lines). The region where 80\,\% of all vetoed surface events are expected is depicted in shaded green. 12 vetoed Pb-recoils (black squares) are observed at $E\sim103$\,keV which can in addition be tagged by pulse shape. The 5 vetoed events at  energies between 130\,keV and 300\,keV are Pb-recoils from the crystal's surface where also part of the energy of the corresponding alpha is absorbed. The areas within red lines at $LY\sim0.22$ ($LY\sim0.01$) indicate the reference region in which degraded alpha ($^{206}$Pb recoil) events would be expected. }
\label{fig:ly_surface_events}       
\end{figure}

First data of TUM40 from the CRESST-II Phase 2 are shown as a LY-energy plot in Fig. \ref{fig:ly_surface_events}. Events in the populated beta/gamma band at $LY\sim1$ are indicated by black dots.  Within the blue lines, namely the 90\,\% upper bound of O recoils and the 90\,\% lower bound for W recoils, at $LY\sim0.1$ the nuclear recoil region is shown. The region where 80\% of all vetoed surface events are expected is depicted in shaded green. Therein,  12 events appear at recoil energies of $\sim$103\,keV which can be  identified clearly as Pb-recoils from $^{210}$Po decays. They are vetoed by   the additional light signal  produced by the  alpha interacting in the polymeric foil. In addition, these events can be tagged by a different light-pulse shape (illustrated by black squares in Fig.  \ref{fig:ly_surface_events}).  This is possible, since the scintillation-time constant of the foil ($\tau\approx100$\,ns) is fast compared to the one of CaWO$_4$ ($\tau\approx500\,\mu$s) \cite{PhysRevB.75.184308}. The events are found at a mean LY of $\sim0.36$ which corresponds to a mean additional light energy of $E_{veto}\approx 37\,$keV$_{ee}$ (in units of electron-equivalent energy). Taking into account that for TUM40 $\sim$1.6\,\% of the deposited energy in the CaWO$_4$ crystal is detected as scintillation light (see section \ref{sec:crystal}), the actual amount of light energy detected from  a 5.3\,MeV alpha impinging the foil is $\sim$0.59\,keV.

The vetoed events at $\sim103$\,keV must originate from $^{210}$Po decays close to surfaces of either the crystal or the surrounding foil. Only the recoiling $^{206}$Pb-nucleus is detected in the phonon channel (the alpha interacts in the foil). Consequently, for geometrical reasons a similar number of events where only the corresponding alpha is detected in the crystal should be observed. In the alpha band, a peak  arises at a  energy of $\sim5.3$\,MeV. Nine events are identified  as alphas from $^{210}$Po \cite{strauss:2014part2} which is consistent with the 12 Pb recoils observed.	
  
In the green band of vetoed events additional 5 events are observed between 130\,keV and 300\,keV. Most probably those originate from shallow implanted $^{210}$Po decays in the  surface of the crystal. Surface events vetoed by the CaWO$_4$ sticks are expected to lie at $LY\sim10$ which is far off the ROI.  Due to limited statistics such events could not be verified.

In Fig. \ref{fig:ly_surface_events} two reference regions are defined (within red lines): 1) for degraded alpha events in the alpha band at $LY\sim0.22$ from 40-300\,keV, 2) for $^{206}$Pb recoils at $LY\sim0.01$ from 40-107\,keV. With 29\,kg-days of exposure both reference regions are free of events (the event at $E\sim85$\,keV in reference region 1 is identified as a Pb-recoil and can be rejected). For comparison, assuming the background level to be as observed in the previous run of CRESST-II, in this exposure,  $8.1\pm2.8$ degraded alphas and $6.9\pm2.6$ Pb-recoils would be expected. This proves the high efficiency of the active surface veto resulting from the new detector design.

\section{Conclusions}
The CaWO$_4$ crystal TUM40 operated in the novel detector module has reached significantly reduced background levels. By supporting the target crystal with CaWO$_4$ sticks  a phonon detector with an excellent performance has been realized. A resolution of \linebreak$\sigma{=}(0.090{\pm}0.010)\,$keV (at 2.60\,keV) and a trigger threshold of $\sim0.60$\,keV were achieved \cite{Angloher:2014myn}. Using a CaWO$_4$ crystal grown at the TUM \cite{erb} the intrinsic background level could be significantly reduced: the average rate of low-energy events (1-40\,keV) amounts to \linebreak  3.44\,/(kg\,keV\,day) \cite{strauss:2014part2}. The surface-event veto due to the fully-scintillating housing rejects backgrounds from  surface-alpha decays with high efficiency. With 29\,kg-days of exposure from the ongoing run of  CRESST-II Phase 2 no events related to degraded alphas or Pb recoils are observed, while these were the dominant identified background source during the previous run of CRESST-II.

Using the first data of TUM40, a low-mass WIMP analysis was performed. Limits on the spin-independent WIMP-nucleon cross section were achieved and a new region of parameter space for WIMP masses below \linebreak 3\,GeV/c$^2$ was probed \cite{Angloher:2014myn}.

The performance of TUM40 demonstrates that the phonon-light technique using CaWO$_4$ as target material has great potential for WIMP searches.

\begin{acknowledgements}	
This research was supported by the DFG cluster of excellence: “Origin and Structure of the Universe”, the DFG “Transregio 27: Neutrinos and Beyond”, the “Helmholtz Alliance for Astroparticle Phyiscs”, the “Maier-Leibnitz-\linebreak Laboratorium” (Garching), the Science \& Technology Facilities Council (UK) and by the BMBF: Project 05A11WOC EURECA-XENON. We are grateful to LNGS for their generous support of CRESST, in particular to Marco Guetti for his constant assistance.
\end{acknowledgements}

%

\bibliographystyle{spphys}       
\bibliography{quenching_bibtex}

\end{document}